\documentclass[12pt]{iopart}

\usepackage{graphicx,iopams}

\newcommand{\equ}[1]{equation~(\ref{#1})}

\newcommand{\eeq}{ \end{equation} }
\newcommand{\beq}{ \begin{equation} }
\newcommand{\bea}{\begin{eqnarray}}
\newcommand{\eea}{\end{eqnarray}}

\newcommand{\ga}{\alpha}
\newcommand{\gb}{\beta}
\newcommand{\gl}{  \lambda }

\newcommand{\fig}[1]{figure \ref{#1}}

\newcommand{\bhu}{ \hat{\bf u} }

\newcommand{\bbr}{ {\bf r} }

\newcommand{\bbv}{ {\bf v } }

\begin{document}

\title{Emergent states in dense systems of active rods: from swarming to turbulence}

\author{H. H. Wensink$^{1,2}$ and H. L\"{o}wen$^1$}

\address{$^1$ Institut f\"ur Theoretische Physik II: Weiche Materie,
Heinrich-Heine-Universit\"at-D\"{u}sseldorf,
Universit{\"a}tsstra{\ss}e 1,  D-40225 D\"{u}sseldorf,
Germany}

\address{$^2$ Laboratoire de Physique des Solides, Universit\'{e} Paris-Sud 11, B\^{a}timent 510, 91405 Orsay Cedex, France}

 \ead{hlowen@thphy.uni-duesseldorf.de}

\date{\today}

\begin{abstract}
Dense suspensions of self-propelled rod-like particles exhibit a fascinating variety
of non-equilibrium phenomena. By means of computer simulations of a minimal
model for rigid self-propelled colloidal rods with variable shape we explore the generic diagram of emerging states
over a large range of rod densities and aspect ratios. The dynamics is studied using a simple numerical scheme for the overdamped noiseless frictional dynamics of a many-body system in which steric forces are dominant over hydrodynamic ones.  
The different emergent states are
identified by various characteristic correlation functions and suitable order parameter fields. At low density and aspect ratio,
a disordered phase with no coherent motion precedes a highly-cooperative swarming state at large aspect ratio.
Conversely, at high densities weakly anisometric particles show a distinct jamming transition whereas slender particles form dynamic laning patterns.
In between there is a large window corresponding to strongly vortical, turbulent flow. The different dynamical states  should be verifiable in systems of swimming bacteria  and artificial rod-like micro-swimmers.
\end{abstract}

\submitto{J. Phys.: Condensed Matter}

\maketitle

\section{Introduction}

Collections of swimming microorganisms and self-propelled particles are able to form remarkable macroscopic
patterns~\cite{2009CoWe,2011KochSub,2009SoAr,2011DrescherEtAl} including swarms~\cite{2010Kearns,2010Ramaswamy}
and complex vortices~\cite{2000Be,2004DoEtAl,2007SoEtAl,2005Riedel_Science,2008SaintillanShelley}.
The tendency for neighbouring particles to align is strongly determined by their mutual interactions which provide the key to
understanding the emergent behaviour at high particle density.  In this regime, the interplay between microscopic self-motility and anisotropic volume-exclusion interactions leads to complex spatio-temporal behaviour that can be directly visualised in two spatial
dimensions, i.e.\ for particles moving in planar confinement. 

Quasi two-dimensional systems of self-propelled particles can be realised 
in a number of ways. Autonomously navigating bacteria and other microbes 
can be confined to free-standing thin films \cite{2007SoEtAl}, between 
solid surfaces \cite{Mino_Clement} or a liquid-gas interface 
\cite{2004DoEtAl, 2011Japan}. On larger length scales, active systems can 
be realised by polar granular rods on a flat vibrating 
surface \cite{kudrolli,2007NaRaMe} or pedestrians moving in complex 
environments \cite{Helbing}. Last not least, colloidal dispersions 
constitute ideal model systems not only for investigating passive matter 
\cite{Lowen2001,Lowen2008} but also for active matter composed of 
self-motile colloidal particles. Over the past decade, a number of 
distinctly different realisations of active colloidal particles have been 
proposed. These include Janus particles driven by catalytic processes 
\cite{Baraban,Bocquet} or thermophoretic \cite{Bechinger} gradients, 
particles propelled by artificial flagella \cite{Bibette} and surface 
waves \cite{snezhko_prl, snezhko_nature} driven in an external magnetic 
field. Rather than being spherical most of these particles have an anisotropic rod-like shape which is found to play a crucial 
role in determining the spatio-temporal behaviour of active particles 
\cite{2006Peruani, Peruani2012}.  Confining systems to quasi-planar 
geometries allows for a direct visualisation of the particles by means of 
real-space microscopy and provides fascinating opportunities to study the 
single-particle and collective behaviour of micro-swimmers.

In this paper we use computer simulation to study a simple model for 
suspensions of rigid, self-propelled rods (SPR) that interact via a 
Yukawa-segment potential \cite{Kirchhoff1996,Wensink2008}. The  
potential allows for a realistic description of the strong mutual 
short-range repulsion that prevents particles from overlapping. 
Self-motility is imposed by introducing a constant propulsion force along the main 
orientation axis of each rod. Consequently, when two neighbouring active 
rods collide they align and the aligning force plays an essential role in 
the formation of flocks of coherently moving particles \cite{vicsek_prl}.  In 
our study we focus on the collective behaviour of dense suspensions of 
strongly interacting particles and characterise the emergent states by 
analysing different correlation functions as dynamical diagnostics. In 
order to retain a generic framework we consider the overdamped frictional 
dynamics of the many-body system where the equations of motion arise from 
a simple force balance between the Stokesian frictional force, the 
collision force and the active force on each rod. Likewise, the rod 
orientations propagate via a torque balance involving the frictional and 
interaction torque acting on each particle. Other forces due to e.g. 
many-body hydrodynamic interactions or thermal fluctuations exerted by the 
embedding solvent are neglected. This allows us to simplify the 
microscopic equations of motion in such a way that the rod aspect ratio 
and density constitute the main variational parameters of the model. The 
microscopic self-propulsion force can be appropriately scaled out and 
subsumed into an (effective) Yukawa amplitude which only has a weak impact 
on the emergent behaviour.

Despite its simplicity the model is capable of predicting a wealth of different steady states that
 hitherto could not be realised within a single framework. Amongst the various states we identify
 an incoherent, disordered dynamical phase at small particle aspect ratio and a cooperative swarming state at larger rod anisometry as found in a number of  particle-resolved models 
\cite{vicsek_prl,2006Peruani,Wensink2008,yang-gompper}.
At high densities and small aspect ratios, we find a jammed phase with distinct local crystalline order. This state is rather common for passive systems \cite{liu-nagel} but  less obvious for active systems. At large aspect ratio and high density, stratified patterns emerge consisting of lanes driven in opposite directions. These structures  are reminiscent of laning patterns observed for mixtures of passive particles  (i.e.\  with no internal driving force)   driven in a 
macroscopic external field  \cite{Dzubiella2002,RexEPJE,WysockiPRL,Vissers2011}.  A similar phenomenon was unveiled recently in mixtures of active and passive rod-like particles \cite{Baskaran_SoftMatter2012}. 
For intermediate densities and aspect ratios, we find distinct chaotic states characterised  by meso-scale turbulent flow patterns with a significant vorticity in the velocity field \cite{2007Cisneros}. 
This type of active  turbulence has been
observed  in microbial suspensions~\cite{2004DoEtAl,2007SoEtAl,2007Cisneros,2008Wolgemuth}. Contrary to traditional turbulent flow observed at high-Reynolds-number passive fluids the vortices that make up the  turbulent flow patterns have a uniform mesoscopic size irrespective of the density or particle shape.  

In principle, the full variety of different emergent states advanced here should be verifiable for bacterial systems and artificial
rod-like colloidal or granular micro-swimmers. In a recent study, the statistical properties of the turbulent states as predicted from the SPR model have been systematically compared with flow-field data of confined bacterial systems \cite{Wensink_Dunkel}. It would be interesting
to pursue a more systematic comparison  with bacterial systems and assemblies of man-made micro-swimmers in order to verify the full topology of the predicted phase diagram. 

The remainder of this paper is organised as follows: in section 2 we specify our model for self-propelled rods, the corresponding equations of motion and the simulation methodology.
Numerical results on the non-equilibrium phase diagram are presented and analysed in section 3.
We conclude in section 4 with a brief discussion of possible extensions of the model and we highlight opportunities to observe the predicted behaviour in experiment.

\section{Frictional dynamics of a self-propelled-rod (SPR)  model}

\begin{figure}[t]
\centering
\includegraphics[clip=,width= 0.35 \columnwidth]{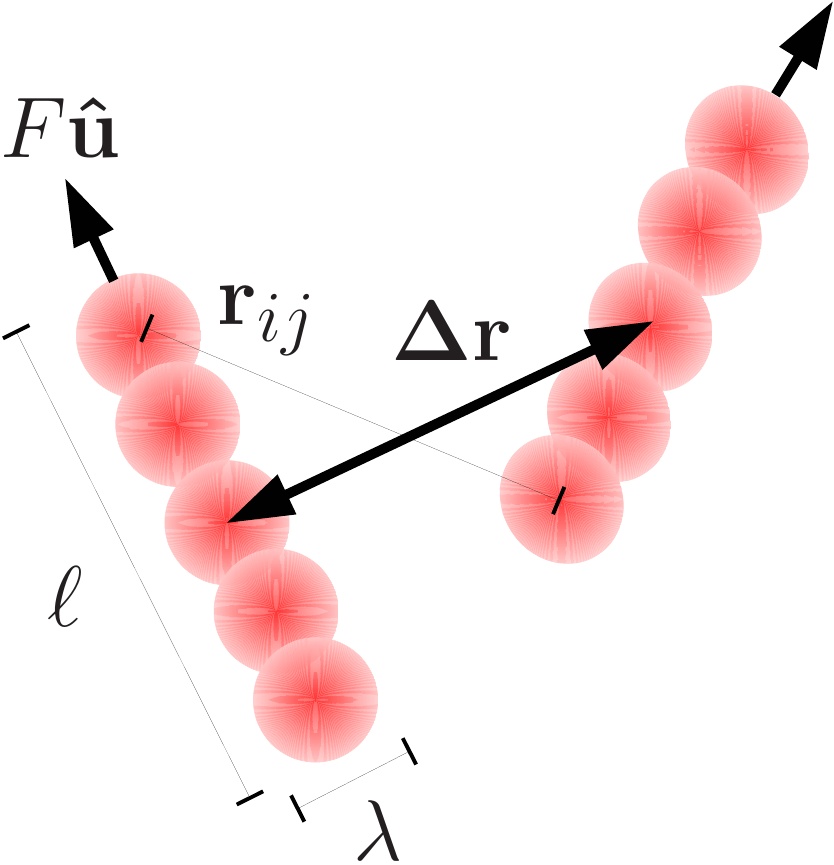}
\caption{Coarse-grained representation of a pair of rod-like micro-swimmers with $n=5$ repulsive Yukawa segments and aspect ratio $a = \ell/\lambda$. Self-propulsion is provided by  a constant  force $F$ acting along the main  rod axis indicated by the orientational unit vector $\bhu$.   The total rod pair potential  is obtained by a sum over all Yukawa segment pairs with distance ${\bf r}_{ij}$ and is a function of the  centre-of-mass distance vector $\Delta \bbr$ and orientations (\equ{upot}). 
 \label{sketch}
}
\end{figure}

One of the simplest ways to envisage a suspension of active mesogens is by considering a collection of rigid,  self-propelled rods  each moving with a constant
  self-motile force $F$ directed along the main rod axis (see \fig{sketch}). Mutual rod repulsion is implemented by discretising each rod into $n$ spherical segments and imposing a
repulsive Yukawa force  with characteristic decay length $\gl$ between the segments of any two rods, such that  $\gl$
defines the effective diameter of the rod of length $\ell$  \cite{Kirchhoff1996}. If two sufficiently long rods perform a
pair-collision, the interaction results in an effective nematic (apolar) alignment  while the centres-of-mass attain a certain minimal distance due to the repulsive Yukawa forces.
The potential energy of a rod-pair $\ga$ and $\gb$ with orientation unit vectors $\{ \bhu_\ga, \bhu _{\gb} \}$ and center-of-mass distance $ \Delta \bbr_{\ga\gb}$,
  is given by 
 \bea
\label{upot}
 U_{\ga\gb} = \frac{U_{0}}{n^2}
 \sum _{i=1}^{n} \sum _{j=1}^{n} \frac{\exp [ - ( r^{\ga\gb}_{ij}/\lambda )  ] }{r^{\ga\gb}_{ij}}.
 \eea 
 where  $U_{0}$ is the potential amplitude, $\lambda $ the
 screening length, and   
 \bea
 r^{\ga\gb}_{ij} = | \Delta \bbr_{\ga\gb}  + (l_{i} \bhu_{\ga} - l_{j}\bhu_{\gb}) |,
 \eea
 the distance between the $i$th segment of rod $\ga$ and the $j$th segment of rod $\gb$, with $l_i\in[-(\ell-\gl)/2,(\ell-\gl)/2]$ denoting the position of segment $i$ along the symmetry axis of
the rod~$\ga$. The screening length $\gl$ defines the effective diameter of the segments such that we may introduce an aspect ratio $ a=  \ell / \lambda $ to quantify the effective anisometry of the SPR.
The case $a = 1$ corresponds to a single
Yukawa point particle ($n=1$). For  $a> 1$, the number of segments per rod is
fixed  as $n=3$ for $1 < a \leq 3 $ and  $n = \lfloor  9 a /8 \rceil $
for $a > 3$ with $ \lfloor \cdot \rceil$ denoting the nearest integer.

We focus on  the dynamical regime relevant to micro-organisms and artificial self-motile colloidal mesogens and we assume the motion of  the
SPRs to be overdamped due to solvent friction (in the zero Reynolds number limit). Since we are interested in the collision-dominated dynamics in dense suspensions, we disregard thermal and intrinsic fluctuations of e.g. bacterial orientation \cite{2011DrescherEtAl}.  Consequently, the equations of motion for the center-of-mass $\bbr_\ga(t)$ and orientation $\bhu_{\ga}(t)$ of each SPR  are entirely deterministic and can be written compactly as 
\bea
\label{e:eom_r}
{\bf f }_{T}  \cdot \partial_{t}\bbr_{\ga} 
&=&  -\nabla_{{\bbr}_{\ga}}  U + F \bhu_{\ga}, 
\\ 
{\bf f}_{R} \cdot \partial_{t} \bhu_{\ga}
&=&  
-\nabla_{\bhu_{\ga}} U.
\label{e:eom_u}
\eea 
Here, $F$ is a constant self-motility force acting along the longitudinal axis of each rod (\fig{sketch}),  $U=(1/2)\sum_{\gb,\ga:\gb\ne\ga} U_{\ga\gb}$ the total
potential energy, $\nabla _{\bhu}$ denotes the gradient on the unit circle, and
\bea
{\bf f}_{T} &=&  
f_0  \,\left[  f_{\parallel} \bhu_{\ga} \bhu_{\ga}  + f_{\perp} ({\bf  I} - \bhu_{\ga} \bhu_{\ga})  \right],\\
{\bf f}_{R} &=& f_0\, f_{R} {\bf I},
\eea 
are the translational and rotational friction tensors ($\bf I$ is the 2D unit tensor) 
with a Stokesian friction coefficient $f_0$. The  dimensionless geometric factors
$\{ f_{\parallel}, f_{\perp}, f_{R} \}$ depend solely on the aspect ratio
$a$, and we adopt  the standard expressions for rod-like macromolecules, as given in Ref.~\cite{tirado}
\bea
\frac{2\pi}{f_{\parallel}} &=&  \ln a - 0.207 + 0.980a^{-1} - 0.133a^{-2},
\\
\frac{4\pi}{f_{\perp}}&=&\ln a+0.839 + 0.185a^{-1} + 0.233a^{-2},
\\
\frac{\pi a^2}{3f_{R}} &=& \ln a - 0.662 +  0.917a^{-1} - 0.050a^{-2}.
\eea
It is expedient to  multiply \equ{e:eom_r} with the inverse matrix ${\bf f}_{T}^{-1}$:
\beq
 \partial_{t} \bbr_{\ga} =  v_{0} \bhu_{\ga} - {\bf f}_{T}^{-1}\cdot \nabla_{{\bbr}_{\ga}}  U,
\eeq
where 
\beq
\label{e:rod_speed}
v_{0}=\frac{F}{f_0 f_{||}},
\eeq
defines the self-propulsion velocity of a non-interacting SPR. 

In our simulations, we have adopted characteristic units such that $\gl =1$, $F=1$, and $f_0=1$, which means 
that distance is measured in units of $\gl$, velocity in units of $F/f_0$, time in units of 
$\tau_{0} =\lambda f_0/F$  and energy in units of $F\gl$. Upon rescaling to dimensionless coordinates,  three relevant system parameters
remain:  The dimensionless  Yukawa amplitude $\tilde{U}_{0} =  U_{0}/(F\lambda)$, which determines
the hardness of the rod interactions relative to their characteristic propulsion energy, the aspect ratio~$a$, and the effective
volume fraction of the system 
\beq
\phi = \frac{N}{A} \left [\gl (\ell - \gl)+\frac{\pi \gl^2}{4} \right ],
\eeq
where the term between brackets denotes the 2D volume $A_{\rm rod}$ of a spherocylindrical  rod. For steeply repulsive Yukawa interactions,  the general dynamical behaviour resembles that of hard rods and does only weakly depend
on the Yukawa amplitude, and we fix $\tilde{U}_{0} = 250$. The remaining quantities, 
the rod shape $a$ and volume fraction $\phi$ constitute the main steering parameters for our investigations.
 We simulate the evolution of the many-body SPR model as a function of time $\tau = t/\tau_{0}$ in a square box of length $L$ with
periodic boundary conditions  at volume fractions in the range
$0.05 < \phi < 0.9$. The simulations  are carried out using a  time discretisation $\Delta \tau =
0.002 \rho^{-1/2}$, where $\rho=N \lambda^{2}/A$ with typically $N =  10^{4}$ and  rods per simulation.  Initial configurations, generated from a rectangular lattice
of aligned rods with $\bhu$
 pointing randomly up or down are allowed to relax during an interval $ \tau =
 1000 $ before statistics is being gathered over an interval $\tau = 20 L$ with $L = (N/\rho)^{1/2}$  the dimension of the
 simulation box (in units of $\lambda$). Velocity vector fields $\bbv (\bbr, t)$  are constructed by measuring the average centre-of-mass velocity within small sub-cells centered around the position $\bbr$. To this end we project the particle  positions onto a 2D cubic grid  $\{(i, j)\;| \;1\le i,j\le G\}$ and measure the average
velocity ${\bf v}(t; i,j)$ in each bin $(i,j)$ at a given time $t$.
In order to test for finite size effects, we consider two different system
sizes:  \lq small\rq\space systems with $N=1 \cdot 10^{4}$ particles and  \lq large\rq\space systems  with $N=4 \cdot 10^{4}$ particles at the same filling fraction $\phi$.  The coarse-graining parameter $G$ is chosen adaptively such as to ensure each bin to represent the average velocity of about 10 SPRs.  Generally, we observe that the dynamical structure and order parameters of the emergent states are robust with respect to changes in the particle number $N$, provided  $N$ is at least of ${\cal O}(10^{4}$). 

\section{Results}

\subsection{Non-equilibrium phase diagram for the SPR model}

\begin{figure}[t]
\centering
\includegraphics[clip=,width= 1 \columnwidth]{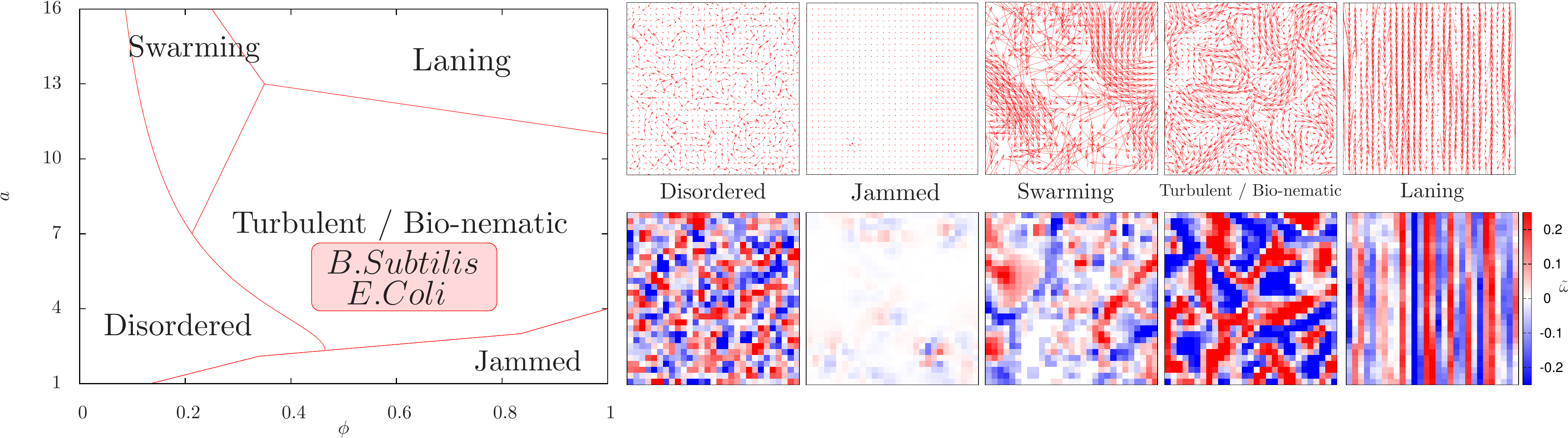}
\caption{
Schematic non-equilibrium phase diagram of the 2D SPR model at variable aspect ratio $a$ and effective filling fraction $\phi$. Values exceeding unity are,  in principle, possible due to  the softness of the Yukawa interactions. The area relevant to self-motile bacteria is highlighted in red.    
 A number of distinctly different dynamical states are discernible as indicated by the coarse-grained maps of the velocity field $\bbv(\bbr,t)$ (upper panels) at time $t$ and the corresponding scalar vorticity field  $\tilde {\omega} (\bbr,t) = [ \nabla \times  \bbv (\bbr,t) ]  \cdot {\bf \hat{e}}_{z}$ (lower panels) expressed in units of $\tau_0^{-1}$. 
 \label{states}
}
\end{figure}

Upon varying the  effective volume filling fraction $\phi$  and the rod aspect ratio~$a$ a number  of qualitatively different
dynamical phases emerge. A schematic non-equilibrium phase diagram, shown in \fig{states}, illustrates the importance of the SPR anisometry in determining the stationary dynamical state of the system.   
The low density regime is generally characterised by  disordered dynamics with little or no cooperative motion. Beyond a certain threshold density  cooperative motion becomes manifest and translates into dynamical states whose structure depends on the intrinsic  `aligning force' of the SPRs. Short rods generally jam at high packing fractions  whilst very long
rods ($a > 13$)   exhibit
swarming  behaviour with large spatio-temporal density
fluctuations. The swarming and laning phases adjoin a large region  of
bio-nematic and turbulent flow characterised by vortices and extended nematic jet-like structures \cite{2007Cisneros,2011Cisneros_PRE}.

Generally, the transitions from the dilute phase to regimes with strong cooperative motion can be localised by the 2D Onsager overlap density~\cite{onsager}, defined as the density  corresponding to a singe rod occupying an average area equal to its excluded area $A_{\rm{ex}}=(2/\pi)(\ell - \lambda)^{2} + (\pi/4)\lambda^{2}$. The latter expression can be derived from the rod dimensions in \fig{sketch} by assuming a pair of spherocylindrical rods with isotropic orientations. By combining terms one arrives at the following expression for the overlap density:
\beq
\phi^{\ast} = \frac{A_{\rm rod}}{A_{\rm ex}} = \frac{1+4(a-1)/\pi}{1+8(a-1)^{2}/\pi^2}.
\eeq
This quantity delimits the density regime beyond which many-body rod collisions (exceeding the pair level) are starting to become important and various non-trivial emergent states arise. In the sections below we shall discuss these in more detail.

\subsection{Short rods: active jamming }

\begin{figure}[t]
\centering
\includegraphics[clip=,width= 1 \columnwidth]{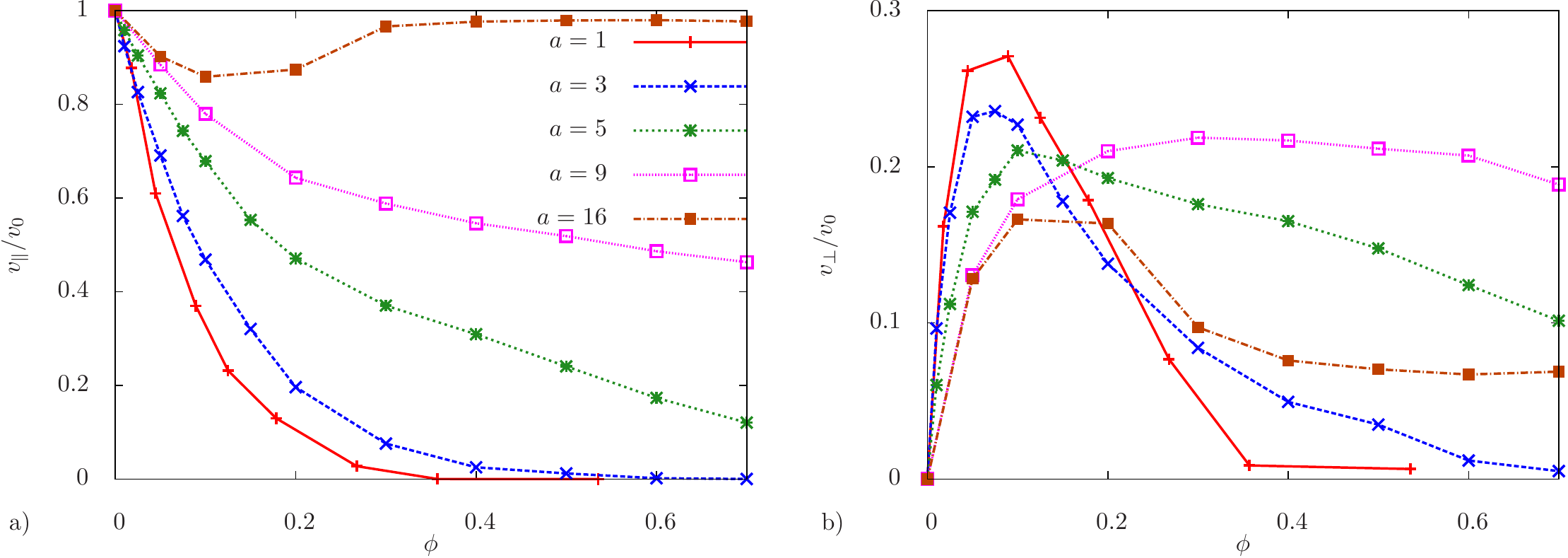}
\caption{Evolution of the average SPR velocity  as a function of filling fraction $\phi$ for a number of particle aspect ratios. Shown is the average velocity component $v_{\parallel}$ along the main rod orientation (a) and the average perpendicular component  $v_{\perp}$ (b), both expressed in units of the velocity $v_{0}$ of a free SPR.
\label{velo}}
\end{figure}

\begin{figure}[t]
\centering
\includegraphics[clip=,width= 1 \columnwidth]{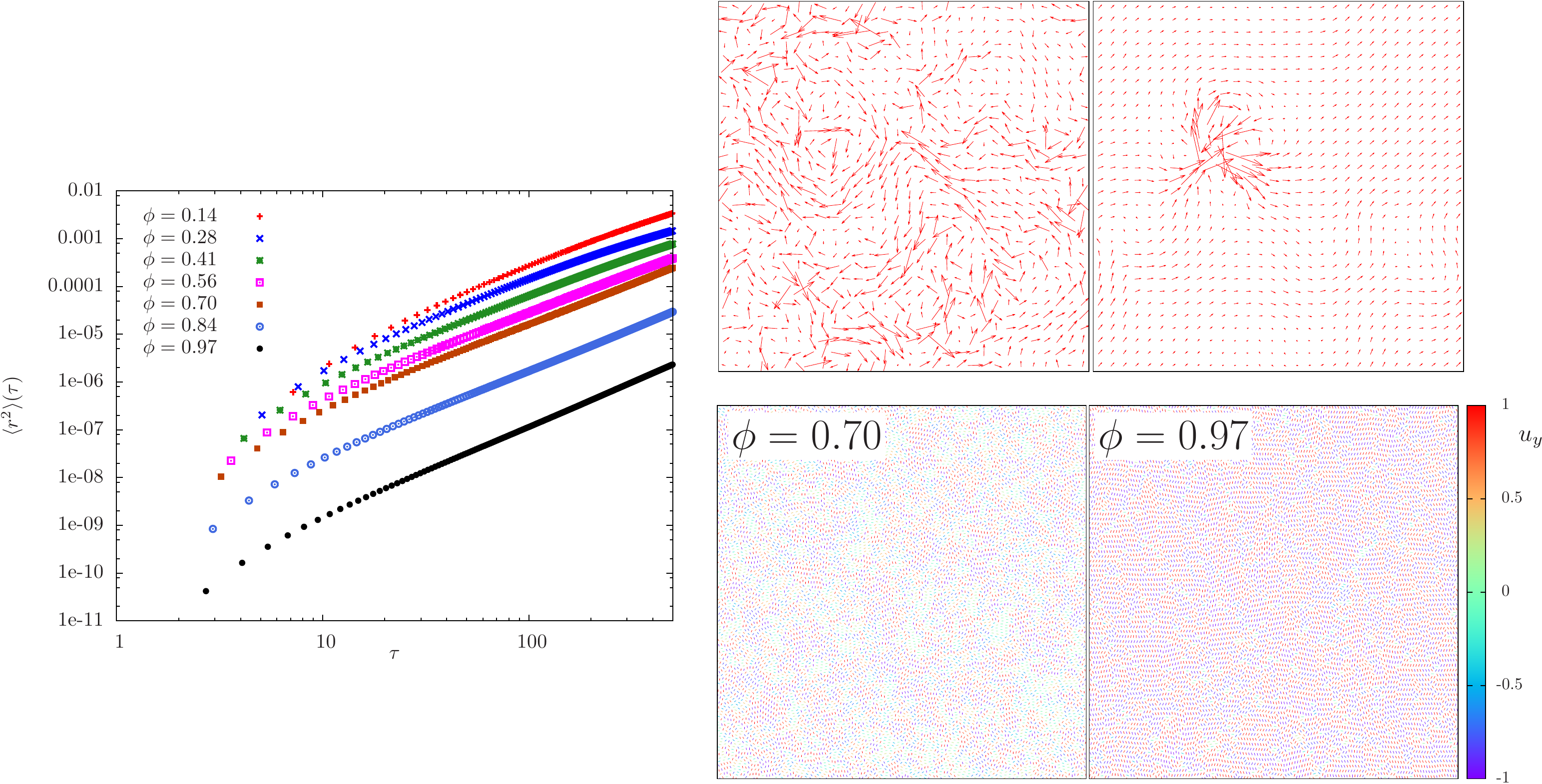}
\caption{Mean-square displacement of the centre-of-mass  for SPRs with aspect ratio $a =3$. The snapshots depict velocity fields (upper panels) and the SPR coordinates (lower panels) for two different bulk filling fractions corresponding to a dynamically disordered fluid at $\phi=0.70$ and a jammed state at $\phi=0.97$. Colour coding is used to indicate the orientation $u_{y} = \bhu \cdot {\bf \hat{e}_{y}}$ of each rod.
\label{msd}}
\end{figure}

For small aspect ratios $a >3 $ a distinct transition towards a
jammed state is observed upon increasing density. This behaviour is
hinted at by the average SPR velocity for which we may probe both parallel and transverse contributions via 
\bea
 v_{\parallel}  & = &  \frac{1}{N} \left \langle \sum_{\ga=1}^{N} \bhu_{\alpha} \bhu_{\alpha}(t) \cdot \bbv_{\alpha}(t) \right \rangle,   \nonumber \\
 v_{\perp}  & = &  \frac{1}{N} \left \langle \sum_{\ga=1}^{N} \left ( {\bf I} -   \bhu_{\alpha} \bhu_{\alpha}(t) \right ) \cdot \bbv_{\alpha}(t) \right \rangle,  
\label{velco}
\eea
where the brackets $\langle \cdots \rangle$ denote a time average. The results are depicted in \fig{velo}. In  general, the average parallel velocity decreases monotonically with density as the particles get progressively hindered in their motion due to mutual rod collisions. For small $a$ the mobility drops rapidly for larger $\phi$ until reaching a threshold level at virtually $v_{\parallel}\sim 0$ indicating dynamical arrest. This behaviour is more clearly reflected in the mean-square
displacement (\fig{msd}) where a sharp drop in the mobility (over nearly 2 orders of
magnitude) at $\phi = 0.84$  marks the onset of
jamming. Throughout the  density range the motion is observed to be sub-ballistic at
long times with $\langle  r^{2} \rangle \sim \tau ^{1.75 \pm 0.1}$.
The jamming point depends strongly on particle anisometry as indicated
in \fig{states} with a marked shift towards higher volume fractions upon increasing $a$. From a structural point of view  the jamming transition is accompanied by a crossover towards   orientationally and positionally ordered structures as evident from the marked degree of local crystalline order at large filling fractions. The velocity maps reveal small pockets of locally enhanced particle mobility that point to the presence of dynamical heterogeneities as commonly found in glassy systems at finite temperature.   A detailed account of freezing and glassy behaviour of self-motile spherical Yukawa particles has been  reported in Ref. \cite{bialke}.

\subsection{Intermediate aspect ratio:  vortical states and turbulence}

\begin{figure}[t]
\centering
\includegraphics[clip=,width= 1 \columnwidth]{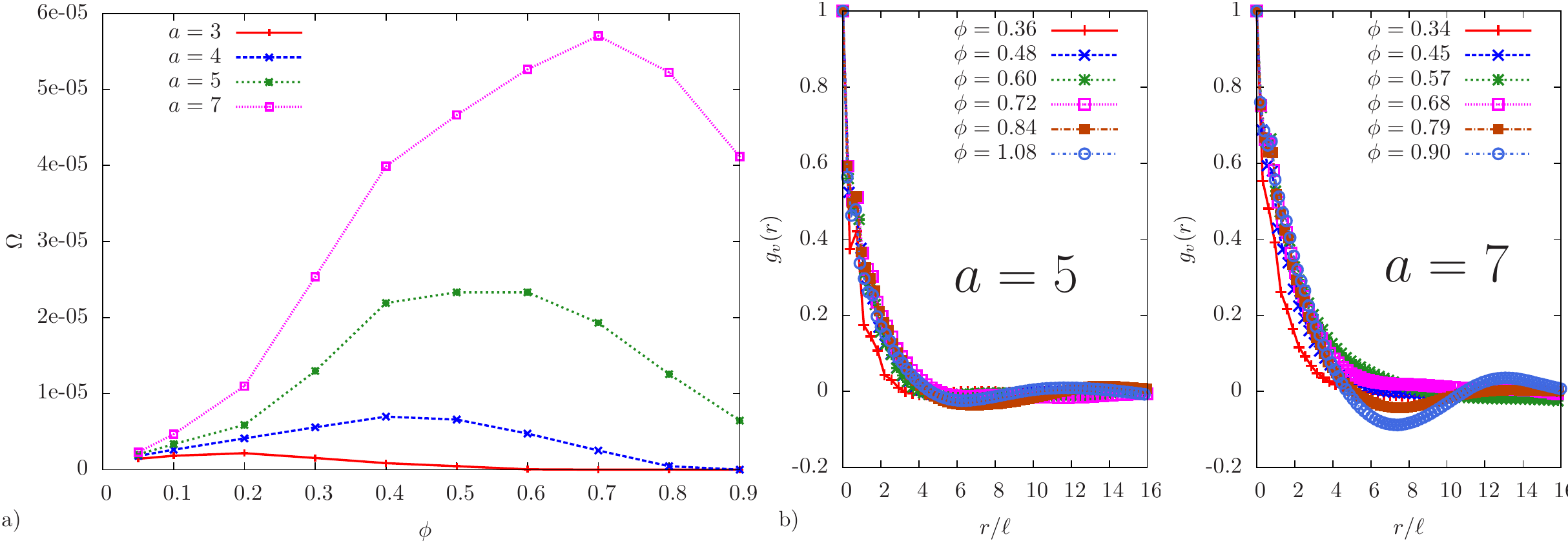}
\caption{(a) Enstrophy $\Omega$ (in units $\tau_{0}^{-2}$) versus filling fraction for a number of aspect ratios $a$ in the turbulent regime. The maxima correspond to the densities where mixing due to vortical motion is the most efficient. (b) Spatial velocity autocorrelation function for a number of bulk volume fractions in the turbulent flow regime for two different aspect ratios $a$.  
\label{turb}}
\end{figure}

\begin{figure}[t]
\centering
\includegraphics[clip=,width= 0.7 \columnwidth]{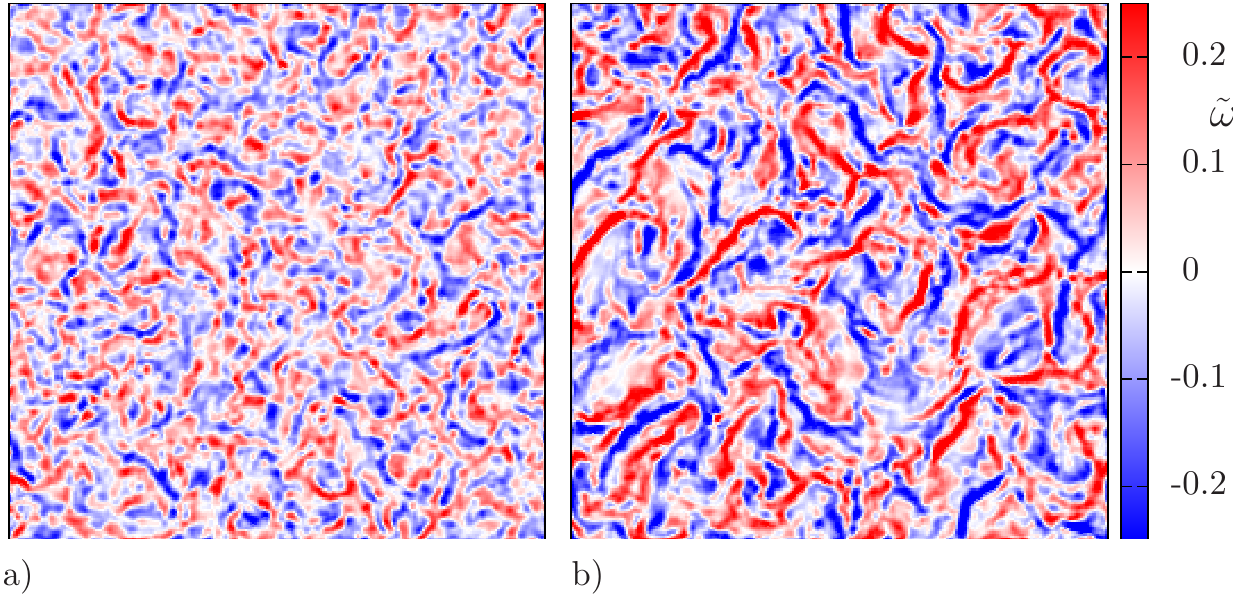}
\caption{Maps of the vorticity field $\tilde {\omega} (\bbr,t) =[ \nabla \times \bbv (\bbr,t)] \cdot {\bf \hat{e}}_{z}$ expressed in units of $\tau_0^{-1}$ showing large-scale turbulent flow for SPRs at intermediate aspect ratios: (a) $\phi = 0.72$, $a=5$ and (b) $\phi = 0.90 $, $a=7$. The snapshots are based on $N=4.10^{4}$ SPRs. The lateral box dimensions are $103 \ell$ (a) and $78 \ell$ (b).    
\label{maps}}
\end{figure}

The maximum in the transverse SPR velocities depicted in \fig{velo}b suggest that the SPRs exhibit some degree of collective swirling motion at moderate densities even at small aspect ratios.  This type of motion becomes much more manifest at larger $a$ where distinct vortical patterns arise akin to turbulent flow.   The kinetic energy
associated with local vortical motion can be measured from the {\em enstrophy} per unit area~\cite{2000Danilov,2002Kellay,2004Frisch}  which is defined as:
\beq
 \Omega  = \frac{1}{2} \left \langle \overline{ |\tilde{\omega}(\bbr,t)|^{2}} \right  \rangle,
\eeq
where the overbar denotes a spatial average. 
For slender rods ($a \geq 3$) the mean enstrophy exhibits a pronounced maximum as a function of the volume fraction~$\phi$ (\fig{turb}b). This maximum signals the density at which vortical motion is maximal. In a bacterial suspension this extremum would correspond to the optimal concentration for fluid mixing. The range of aspect ratios over which turbulence flow is stable
corresponds well with the typical aspect ratios of bacterial cell bodies, e.g. $a\sim 3$  for {\it E. Coli}  and $a\sim 6$ for {\it B. subtilis} ({\em cf.} \fig{states}). 

The typical size of the vortices that make up the turbulent flow patterns  can be extracted 
from the equal-time velocity autocorrelation
function  (VACF) $g_{v}(r) = \langle \bbv(0,t) \cdot \bbv(\bbr , t) \rangle$. This quantity can be obtained from the  microscopic SPR coordinates $\{ \bbr_{\ga}, \bbv_{\ga} \}$ via:
\beq
g_{v}(r) =  \frac { \left \langle \sum_{\ga} \sum _{\gb \neq \ga} \delta (r - |\bbr_{\ga}
    - \bbr_{\gb}|) (\bbv_{\ga} \cdot
  \bbv_{\gb}  - \left \langle v \right \rangle^2 ) \right \rangle }{\left \langle \sum_{\ga}
  \sum_{\gb \neq \ga} \delta (r - |\bbr_{\ga} - \bbr_{\gb}|) ( \left \langle v^{2} \right \rangle  - \left \langle v \right \rangle ^{2} ) \right \rangle }.
\label{vacf}
\eeq
The decay of the VACFs in \fig{turb}b reveals a typical vortex size of about $\sim 5\ell$, an estimate that seems rather insensitive to the bulk density and aspect ratio. Monotonically decreasing velocity correlations correspond to bio-nematic-type states where large-scale nematic jets and vortices coexist \cite{2007Cisneros} whereas negative correlations (cf. the curves for $a=7$ and $\phi>0.8$) represent  more pronounced vortical collective motion reminiscent of fully developed meso-scale turbulent flow \cite{Wensink_Dunkel}. Typical vorticity snapshots  are shown in \fig{maps}.

\begin{figure}[t]
\centering
\includegraphics[clip=,width= 0.4 \columnwidth]{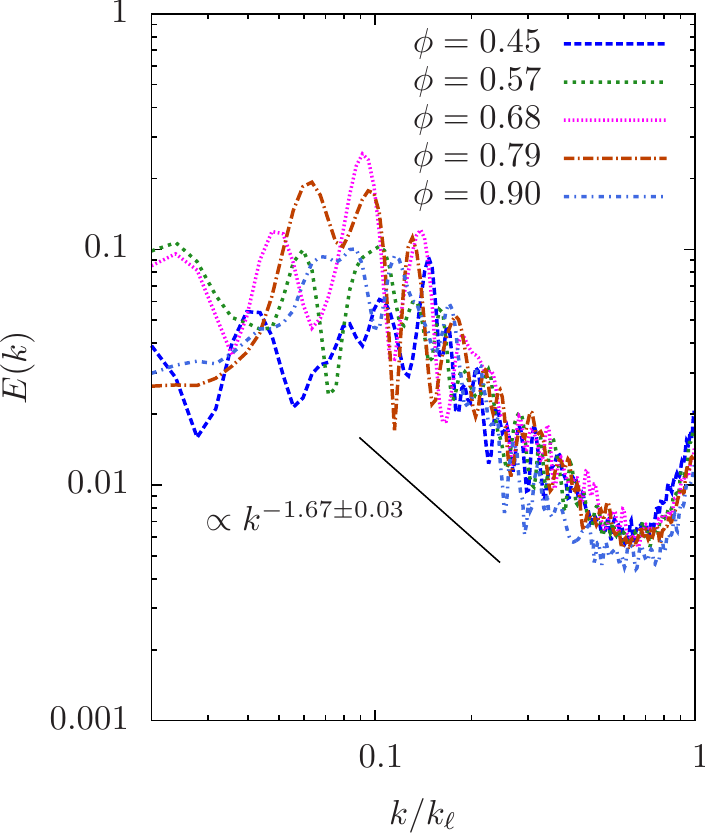}
\caption{Power spectra of the kinetic energy  for turbulent flow of SPRs with $a=7$ ($k_{\ell}=2\pi/\ell$). Universal scaling behaviour (with scaling exponent $-5/3$) is observed  in the  intermediate range of wavenumbers $k$.     
\label{spec}}
\end{figure}

In order to make a connection with classical 2D turbulence in high-Reynolds number fluids we have calculated the energy spectrum which can be obtained as a Fourier transform of the VACF:
\beq
E(k) = \frac{k}{2 \pi} \int d \bbr \exp [ -i {\bf k}  \cdot \bbr ] \langle  \bbv(0,t) \cdot \bbv(\bbr, t) \rangle. \label{ek}
\eeq
An alternative and more formal definition reads  $\langle v^{2} \rangle = 2 \int_{0}^{\infty}dk E(k)$  where $E(k)$ reflects the accumulation of kinetic energy over different length scales.
The results in \fig{spec}  suggest asymptotic power law scaling regimes for intermediate $k$-values with a power-law exponent close to the characteristic $k^{-5/3}$-decay predicted by the Kolmogorov-Kraichnan scaling theory \cite{1967Kraichnan_1,1980Kraichnan} for (passive) 2D turbulence in the inertial regime.  In the present case, however, inertia is absent on the particle scale because the SPR motion is completely overdamped but it is possible that the self-propulsion establishes `effective' collective inertial effects on larger scales which could explain the observed $k^{-5/3}$ decay.  Contrary to regular turbulent flow where energy is injected on the macroscopic scale, active turbulence is characterised by forcing on the microscopic scale.  In general, the transport of kinetic energy towards smaller $k$ becomes significantly damped on larger length scales \cite{2011Japan} as highlighted by the low-$k$ plateau in the power spectra in \fig{spec}.  We refer the reader to Ref. \cite{Wensink_Dunkel} for a more  detailed discussion  comparing meso-scale turbulence in active suspensions and regular high-Reynolds-number turbulent flow.

\subsection{Long rods: swarming and lane formation}

\begin{figure}[t]
\centering
\includegraphics[clip=,width= 1 \columnwidth]{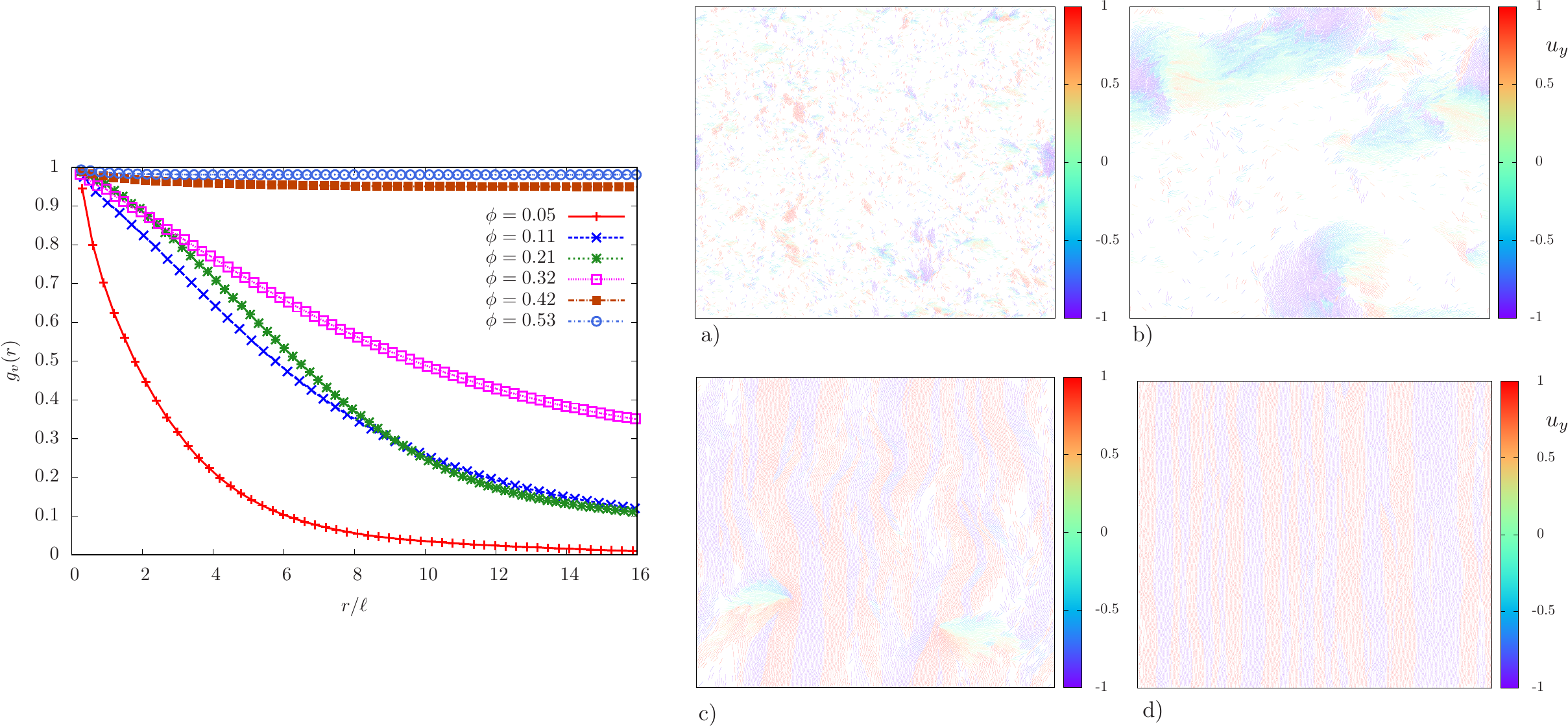}
\caption{Spatial auto-correlation functions of the vertical velocity component $g_{v_{y}}(r)$  (\equ{vacf}) at various filling fractions $\phi$  for SPRs with $a =16$. Particle snapshots of the stationary states at (a) $\phi = 0.05$ - disordered, incoherent motion  (b) $\phi = 0.21$ - swarming, (c) $\phi = 0.32$ - initial stage of laning formation, and (d) $\phi=0.42$ - fully developed laning state. The lateral box dimension corresponds to $38 \ell$.  Colour coding is used to indicate the orientation $u_{y} = \bhu \cdot {\bf \hat{e}_{y}}$ of each rod.
\label{lan}}
\end{figure}

At low to moderate density slender rods with $a >10$  tend to form large compact flocks
reminiscent of the cooperative motion observed in large groups of organisms, e.g.  fish schools, bird flocks (see \fig{lan}a-b) \cite{1998TonerTu_PRE, 2005ToTuRa}.
At larger volume fractions distinct laning patterns emerge that consist of cooperative stratified motion (\fig{lan}c-d). The transition from flocking to laning  as a function
of the volume fraction can be localised from the equal-time VACF for the velocity components along the lane directions (in this case the vertical $y-$component). The characteristic decay of the VACF $g_{v_{y}}(r)$,  shown in
\fig{lan}, allows a distinction between the disordered state at low densities where rod clusters are small and velocity correlations decay rapidly  and the emergence of large cooperative flocks with velocity correlations spanning several dozens of rod lengths.  A marked divergence of the correlation length occurs around
$\phi \approx 0.4 $ where the
flocks start to span the entire system and self-organise into lanes moving in opposite
directions.  The laned patterns remain stable throughout the sampled time interval and no sign of break-up is observed over time even for large systems. We have verified  the stability  of the lanes against small thermal fluctuations of the rod orientations that could be induced by the embedding medium or by some internal source, e.g., bacterial flagella. The rotational fluctuations are represented by a Gaussian white noise contribution $\Delta \bhu$ to the equation of rotational motion of each rod $\ga$ (cf. \equ{e:eom_u}):  
\beq
\partial_{t} \bhu_{\ga}=  - {\bf f}_{R} ^{-1} \cdot \nabla_{\bhu_{\ga}} U + \Delta \bhu_{\ga}.
\eeq
The stochastic term has zero mean $\langle \Delta \hat{u}_{i \ga} \rangle =0 $
and correlations $\langle \Delta \hat{u}_{i\ga} (t) \Delta
\hat{u}_{j\gb}(t^{\prime}) \rangle = 2 D_{R}^{\ast} \delta_{ij} \delta_{\ga \gb}
\delta(t - t^{\prime})$ (with $i=x,y$) in terms of some effective rotational diffusion rate $D_{R}^{\ast}$. Although `run-and-tumble'  motion as commonly observed in
bacterial systems (notably {\it E. Coli} \cite{2011DrescherEtAl,
tailleurcates}) is strictly non-Brownian at short times, its long-time behaviour  is well-captured  by a rotational diffusion process with a diffusion constant much larger than the Stokes-Einstein value $D_{R} = k_{B}T/f_{0} f_{R}$ (where $k_{B}T$ is the thermal energy) for passive Brownian rods \cite{tailleurcates}. The strength of the tumbling motion is conveniently
expressed in terms of the dimensionless tumbling parameter $\ell
D_{R}^{\ast}/v_{0}$ which is the ratio of the translation time a free SPR needs
to swim over a distance $\ell$ and the typical tumbling time $1/D_{R}^{\ast}$. 
Typical values  for {\it E. Coli} and other
swimming bacteria are $\ell D_{R}^{\ast}/v_{0} \sim 0.01$ \cite{2011DrescherEtAl}. In the dense regime, the
particle velocities are dominated by rod-rod collisions rather than thermal
fluctuations and the intrinsic rotational diffusivity of the SPRs does not
incur any qualitative change to the laning structures.  In general, we
assume that the spatio-temporal states and the topology of the phase
diagram is robust against weak fluctuations in the swimming
direction of the SPRs. We remark that similar laning states were
encountered at finite temperature in binary mixtures of SPRs with
different self-motility \cite{Baskaran_SoftMatter2012}.  In both classes
of driven systems laning instabilities occur if the disparity between the
species mobility exceeds a certain threshold. In our case, however, such
an intrinsic driving force is absent since all particle have equal
mobility.

\section{Conclusions}

We have studied the collective dynamical behaviour of a simple two-dimensional model of self-propelled rigid rods (SPR) by means of numerical simulation. Depending on the rod shape and density, the SPR model exhibits a wealth of different emergent dynamical states including
swarming, turbulence, laning and jamming. Although many of these states have been encountered in various set-ups, most notably (mixtures of) spherical particles in different external fields, the SPR model is able to generate these dynamical states upon variation of only two basic system parameters; the particle shape and density. The present approach may therefore serve as a benchmark to characterise the collective properties of different classes of self-motile organisms and artificial micro-swimmers of various shapes. As for the turbulent state, it was recently shown that the SPR model is capable of reproducing the velocity statistics obtained from experiments on strongly confined {\it Bacillus subtilis\/} suspensions \cite{Wensink_Dunkel}.  Future experiments  on dense systems of self-propelled particles with  low and high particle anisometry will hopefully allow for similar comparisons for the jammed and laned state, respectively.

Future  efforts could be aimed at extending the SPR model and the associated equations of motion by accounting for e.g. multi-body hydrodynamic interactions mediated by the solvent, particle flexibility and body forces transmitted by chemical gradients (chemotaxis) that could be relevant in concentrated bacterial systems. The influence of thermal fluctuations could be incorporated  if one wishes to assess the effect of translation and rotational noise (bacterial tumbling) in more detail.  It is also desirable to explore the SPR model in three spatial dimensions, for instance, to study the phenomenology of fully developed 3D meso-scale active turbulent flow which has been unexplored so far. Finally, it would be challenging to construct microscopic theories that are capable of predicting the observed emergent states advanced in this study. Dynamical density functional theory for anisotropic particles \cite{Wensink2008,RexWensink2007,Wittkowski2012} could provide a promising avenue for this.

\ack

We  are grateful to J\"orn Dunkel and Lyderic Bocquet for helpful discussions. Financial support from the DFG within SFB TR6 (project D3) is
gratefully acknowledged.

\section*{References}
\bibliographystyle{iopart-num}
\bibliography{refs}

\end{document}